\documentclass[11pt]{article}
\usepackage{graphicx}

\newcommand{\BABARPubYear}    {01}

\newcommand{\BABARConfNumber} {27}

\input pubboard/babarsym

\def\Dspstar{\ensuremath{D_s^{*+}}\xspace}
\def\Dsps{\ensuremath{D_s^{(*)+}}\xspace}
\def\Dsphipi{\ensuremath{\Ds \rightarrow \phi \pip}\xspace}
\def\Dsgamma{\ensuremath{\Dspstar \rightarrow \Ds \gamma}\xspace}

\long\def\inst#1{\par\nobreak\kern 4pt\nobreak
    {\it #1}\par\vskip 10pt plus 3pt minus 3pt}

\begin{document}
{\pagestyle{empty}

\begin{flushright}
\babar-CONF-\BABARPubYear/\BABARConfNumber \\
%\babar-PUB-\BABARPubYear/\BABARPubNumber \\
%SLAC-PUB-\SLACPubNumber \\
%hep-ex/\LANLNumber \\
July, 2001 \\
\end{flushright}

\par\vskip 4cm

% Title of the paper
\begin{center}
\Large \bf \boldmath Measurement of \Ds and \Dspstar production in
\B meson decays and from continuum \epem annihilations at 
$\sqrt{s}=10.6\gev$
\end{center}
\bigskip

\begin{center}
\large The \babar\ Collaboration\\
\mbox{ }\\
July 20, 2001
\end{center}
\bigskip

% Abstract
\begin{center}
\large \bf Abstract
\end{center}
New precise measurements of \Ds and \Dspstar meson production 
from \B mesons and \qqbar continuum events near the \FourS resonance
are presented in this paper. Using the \babar\ data recorded in 1999 and 2000
of 20.8\invfb on-resonance and 2.6\invfb off-resonance,
we measure the inclusive branching fractions
$\BR(B\rightarrow \Ds X) = (10.93\pm0.19\pm0.58\pm2.73)\%$ and 
$\BR(B\rightarrow \Dspstar X) = (7.94\pm0.82\pm0.72\pm1.99)\%$,
%$\BR(B\rightarrow \Ds X) = (10.9\pm0.2\pm0.6\pm2.7)\%$ and 
%$\BR(B\rightarrow \Dspstar X) = (7.9\pm0.8\pm0.7\pm2.0)\%$,
where the first error is statistical, the second is the systematic error,
and the third is the error due to the \Dsphipi branching fraction uncertainty.
The branching fractions 
$\Sigma\BR(B\rightarrow \Dsps \Dbar^{(*)}) = (5.07\pm0.09\pm0.34\pm1.27)\%$
and 
$\Sigma\BR(B\rightarrow \Dspstar \Dbar^{(*)}) = (4.07\pm0.42\pm0.53\pm1.02)\%$
%$\Sigma\BR(B\rightarrow \Dsps \Dbar^{(*)}) = (5.1\pm0.1\pm0.3\pm1.3)\%$
%and
%$\Sigma\BR(B\rightarrow \Dspstar \Dbar^{(*)}) = (4.1\pm0.4\pm0.5\pm1.0)\%$
have been determined from 
the measured \Dsps momentum spectra.

\vfill
\begin{center}
Submitted to the\\ 20$^{th}$ International Symposium 
on Lepton and Photon Interactions at High Energies, \\
7/23---7/28/2001, Rome, Italy
\end{center}

\vspace{1.0cm}
\begin{center}
{\em Stanford Linear Accelerator Center, Stanford University, 
Stanford, CA 94309} \\ \vspace{0.1cm}\hrule\vspace{0.1cm}
Work supported in part by Department of Energy contract DE-AC03-76SF00515.
\end{center}
}
\newpage

% Input author list file
\begin{center}
\small

The \babar\ Collaboration,
\bigskip

%% author list as of 12-Jul-2001 (622 authors)
B.~Aubert,
D.~Boutigny,
J.-M.~Gaillard,
A.~Hicheur,
%A.~Jeremie, per J.P.Lees
Y.~Karyotakis,
J.~P.~Lees,
P.~Robbe,
V.~Tisserand
\inst{Laboratoire de Physique des Particules, F-74941 Annecy-le-Vieux, France }
A.~Palano
\inst{Universit\`a di Bari, Dipartimento di Fisica and INFN, I-70126 Bari, Italy }
G.~P.~Chen,
J.~C.~Chen,
N.~D.~Qi,
G.~Rong,
P.~Wang,
Y.~S.~Zhu
\inst{Institute of High Energy Physics, Beijing 100039, China }
% ...

\end{center}\newpage

% The body of the paper starts here

% reset footnote counter
\setcounter{footnote}{0}

\section{Introduction}
\par

\begin{figure}[hb]
\begin{center}
        \includegraphics[angle=-90,width=0.5\textwidth]{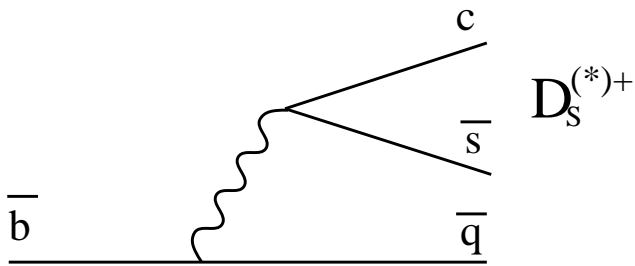}
\caption{
The main spectator diagram leading to the production of \Dsps mesons
in \B decays.}
\label{fig:diagram} 
\end{center}
\end{figure}

The study of \Dsps meson production in \B decays allows exploration of
the mechanisms leading to the creation of $c\overline{s}$ quark pairs.
Although several Feynman diagrams could lead to \Dsps mesons in \B decays,
the spectator diagram (Fig.~\ref{fig:diagram}) is expected to dominate. 
In addition, \Dsps mesons could be produced from \ccbar continuum events.
It has been pointed out~\cite{FWD:1} that the rate from \B decays
may be large. 
This might help to explain some of the theoretical difficulties~\cite{bigi:1} 
in accounting simultaneously for the total inclusive \B decay rate and the 
semileptonic branching fraction of the \B meson. 
The measurement of the \Dsps momentum allows a
determination of the fraction of two-body and multi-body decay modes,
which will aid understanding $b\to c\overline{c}s$ transitions.

In this paper, measurements of $B\rightarrow \Ds X$ 
and $B\rightarrow \Dspstar X$
production rates and momentum spectra$^1$ 
\footnotetext[1]{Reference in this paper to a specific decay 
channel or state
also implies the charge conjugated decay or state. The notation \Dsps means
either \Ds or \Dspstar. $B\rightarrow \Dsps \Dbar^{(*)}$ 
is a general representation of any of 
modes with $c\overline{s}$, $\overline{c}q$ including their excited states.}
are presented. These mesons are reconstructed using the decays \Dsphipi and
\Dsgamma.

\section{The \babar\ detector and data set}
\par
The data used for this analysis
were collected with the \babar\ detector~\cite{babar} 
at the \pep2 asymmetric-energy collider~\cite{pep} at the Stanford 
Linear Accelerator center.
%The results presented here are based on data taken in the 1999-2000 run.
An integrated luminosity of 20.8\invfb was recorded corresponding
to about 22.7 million produced \BB pairs at
the \FourS resonance (``on-resonance'') and 2.6\invfb
at an energy about 40\mev\ below the \BB threshold 
(``off-resonance''). 
Since a detailed description of the \babar\ detector 
is presented in Ref.~\cite{babar}, only the components of the detector
most crucial to this analysis are briefly summarized below.
 
Charged particles are detected
and their momenta measured by a
combination of a central drift chamber (DCH) with a helium-based
gas and a five-layer (double-sided) silicon vertex
tracker (SVT), within a 1.5 T solenoidal field produced by a
superconducting magnet. 
The tracking system covers a solid angle of 92\% in the center-of-mass
frame.
%The charged particle momentum resolution
%is approximately $(\delta p_T/p_T)^2 = (0.0015 \, p_T)^2 + (0.005)^2$,
%where $p_T$ is in GeV$/c$. The SVT, with typically 10\mum single-hit
%resolution, provides impact parameters in both the transverse plane
%and in $z$.
Charged particles are identified using
the ionization energy loss (\dedx) measured in the DCH and SVT
and the Cherenkov radiation detected in a ring imaging Cherenkov 
device (DIRC). Photons are identified by the CsI
electromagnetic calorimeter. 

%\par
%All charged tracks are required to originate from within $\pm$10~cm 
%along the beam direction and $\pm$1.5~cm in the transverse plane of
%the interaction point and leave at least 12 hits in the Drift Chamber.

\section{\boldmath \Ds and \Dspstar selection}
\par
The analysis reported
here uses only the decay mode \Dsphipi, with $\phi \rightarrow { K^+ K^-}$,
as this channel offers the best signal-to-background ratio.
The charged tracks are required to originate from within $\pm$10~cm 
along the beam direction and $\pm$1.5~cm in the transverse plane of
the interaction point and leave at least 12 hits in the drift chamber.

In order to obtain a
sufficiently clean sample, kaon identification is required 
for the tracks forming the $\phi$ meson by
using \dedx information from DCH and SVT
and the Cherenkov angle and the number of photons as measured by the DIRC.
The kaon selection is based on the likelihoods given
by each detector and uses, for each track, the ratio of
likelihoods for the pion and the kaon mass hypotheses, $L_\pi/L_K$.
If this ratio is less than unity for at least one of the considered
subsystems, the particle is selected as a kaon.
The DIRC is used both in a positive identification mode and 
also in a veto mode for the case where a kaon with the measured track momentum
would not be above the Cherenkov threshold.
A tighter level of identification is also available using
a total likelihood defined
as the product of the likelihoods of each subsystem.
In this case the track is tagged as a kaon if the ratio of the total
likelihoods for the pion and kaon mass hypotheses is less than unity.
 
Three charged tracks originating from a common vertex
are combined to form a \Ds candidate. 
Two oppositely charged tracks have to be identified as
kaons by satisfying the basic criteria and at least one of them has
to satisfy the tighter selection.
The ${K^+K^-}$ invariant mass must be within 8\mevcc
of the nominal $\phi$ mass~\cite{pdg}.
In this particular decay, the $\phi$ meson is polarized
longitudinally and therefore the angular distribution of the kaons has a
$\cos^2\theta_{H}$ dependence, where
$\theta_{H}$ is the angle between the $K^+$
and \Ds in the $\phi$ rest frame.
This angle is required to satisfy
$|\cos\theta_{H}|>0.3$, thereby keeping 97.5\% of the
signal while rejecting about 30\% of the background.
 
\begin{center}
\begin{figure}[t]
\begin{minipage}{.48\textwidth}
        \includegraphics[width=\textwidth]{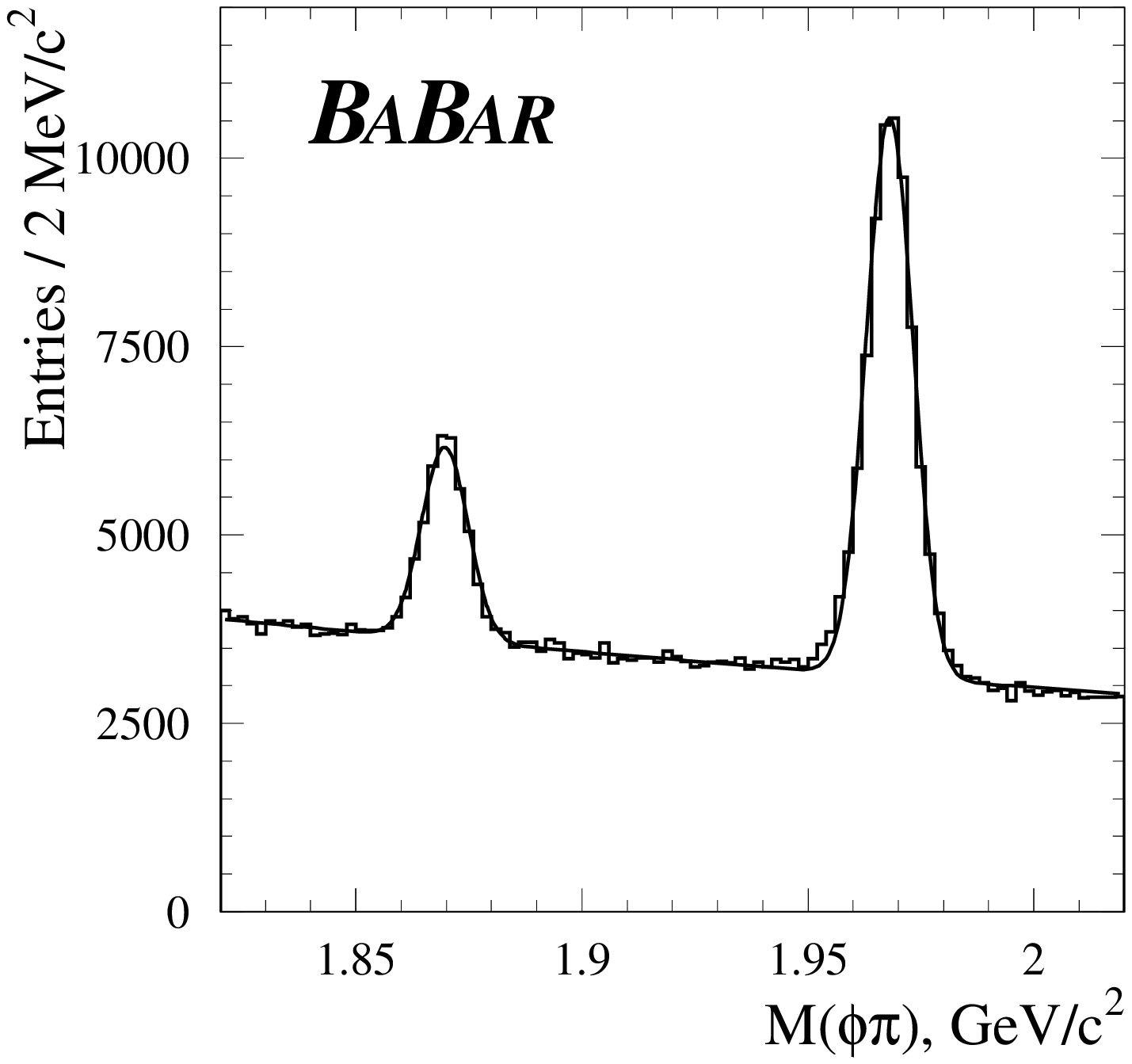}
        \caption[.]
        {
        The observed
        $\phi\pi$ invariant mass spectrum. The lower mass peak corresponds 
        to the Cabibbo-suppressed decay mode $\Dp \rightarrow \phi \pip$.
        The fit function is a single Gaussian for each peak, with their widths
        constrained to be equal, on top of an exponential background.
        }
        \label{fig:ds}
\end{minipage}
\hfill
\begin{minipage}{.48\textwidth}
        \includegraphics[width=\textwidth]{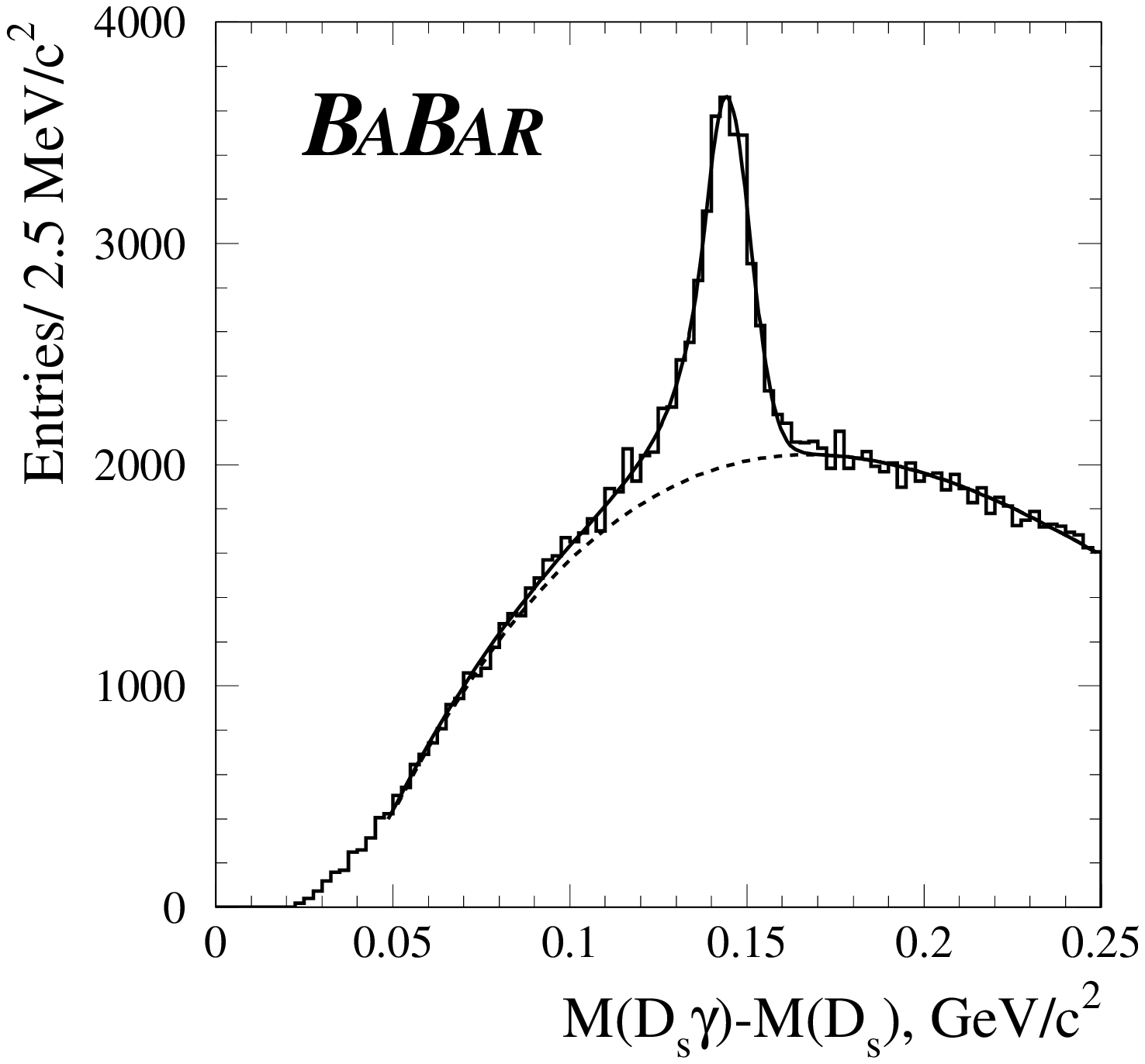}
        \caption[.]
        {Distribution of the
        $\Delta M = M_{\Ds\gamma} -  M_{\Ds}$ 
        mass difference. The fit function is a Crystal Ball function
        for the signal on top of a threshold function, as described in text.
        }
        \label{fig:dsst}
\end{minipage}
\end{figure}
\end{center}

Using the selection described above, a clean \Ds signal
of $47794 \pm 311$ events is observed (Fig.~\ref{fig:ds}). 
A clear signal for the
Cabibbo-suppressed decay mode $\Dp \rightarrow \phi \pip$ 
is also observed.

\Dspstar mesons are reconstructed using the decay \Dsgamma,
with the subsequent decay \Dsphipi.
\Ds candidates are selected by requiring the $\phi\pi$ invariant mass to
be within 2.5 standard deviations ($\sigma$) of the peak value. 
These are then combined with the ``single photons'' of the event,
which are required to satisfy
$E_\gamma > 50$\mev, where $E_{\gamma}$
is the photon energy in the laboratory frame, and
$E_\gamma^* > 110$\mev, where $E^*_{\gamma}$
is the photon energy in the \FourS rest frame.
In order to reduce the combinatoric background,
the candidate photon should not form a $\pi^0$, defined by
a total energy $E_{\gamma\gamma}^* > 200$\mev and
an invariant mass $115 < M_{\gamma\gamma} < 155$\mevcc,
when combined with any other photon in the event. 
The distribution of the mass difference
$\Delta M = M_{\Ds \gamma} - M_{\Ds}$
is shown in Fig.~\ref{fig:dsst}.
A clear peak with $14392\pm 376$ events is observed.

\section{\boldmath Extraction of the \Dsps momentum spectra}
\par
The momentum spectrum of \Ds mesons in the \FourS rest frame is extracted 
by fitting the $\phi\pi$ invariant mass distribution in each momentum
bin. The bins are chosen to be 200\mevc wide,
which is much larger than the momentum resolution ($\approx 6$\mevc).
The fit function is a single Gaussian for each of the \Ds and the \Dp signals, 
with the constraint of a common width.
The combinatorial background is accounted 
for by an exponential. As there are many more events
for the on-resonance data, the number of \Ds in 
the off-resonance data is extracted with 
the same fit function but with 
$M_{\Dp}$, $M_{\Ds}$ and $\sigma$ fixed to the values obtained
from the binned chi-squared fit to the on-resonance data.

In the same way as for \Ds, the momentum spectrum of \Dspstar
mesons in the \FourS rest frame is extracted by 
fitting the $\Delta M$
invariant mass distribution for 250\mevc wide momentum bins.
The $\Delta M$ distribution for the signal 
is characterized by an asymmetric shape to account for
energy leakage and shower shape fluctuations.
The fit function for the signal is the Crystal Ball function~\cite{CBfunc}.
%Gaussian-with-power-law-tail 
For the background, a threshold function 
$$ f(\Delta M) = p_1 (\Delta M - p_2)^{p_3} e^{p_4(\Delta M-p_2)}$$
is used, with the four free parameters $p_i$ being determined from the fit.
%Is $p_1$ simply a normalisation; if so, is it really a fit parameter???
After ensuring that the connection point between the
Gaussian and power-law tail of the Crystal Ball function
does not depend on momentum
and agrees with the Monte Carlo, this parameter 
has been fixed at 0.89$\sigma$ in the final fit. 
The off-resonance data are again fit
with the signal shape parameters $\Delta M$ and $\sigma$
fixed to the values obtained from the fit to the on-resonance data. 

%================================
%The $\Delta M$ distributions for the low momentum of \Dspstar is peaked
%at low masses. This leads to some difficulties for the extraction of the
%number of events, specially for off-resonance data. We impose additional 
%systematic error for the first points to the knowledge of the momentum 
%spectrum obtained as a variation of the number of events with different 
%parameterizations of background.
The uncertainty on the shape of the background leads to an additional 
systematic error.
This error is estimated by using different parameterizations for the
background shape.
%================================

\begin{center}
\begin{figure}
\begin{minipage}{.48\textwidth}
        \includegraphics[width=\textwidth]{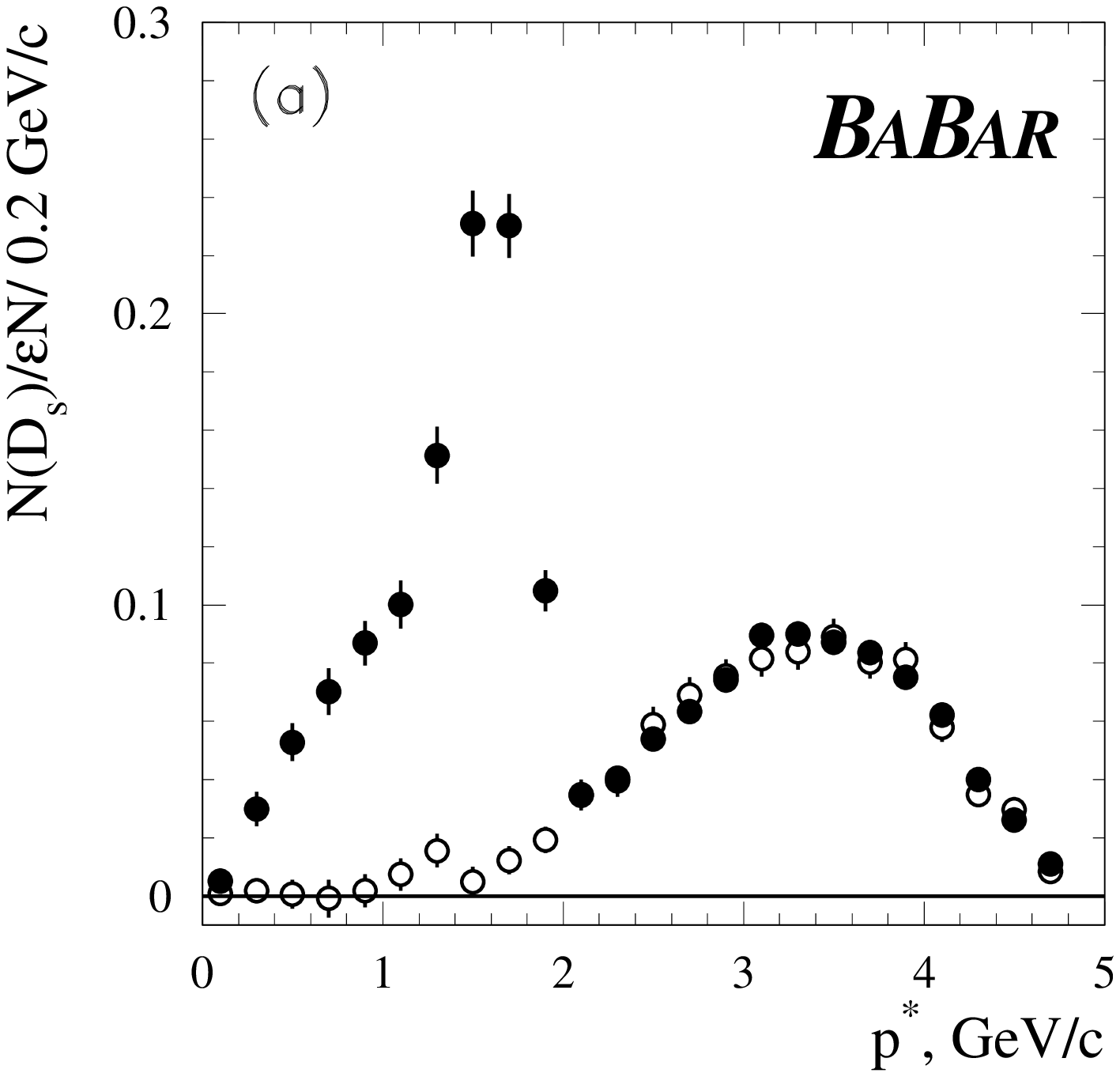}
\end{minipage}
\hfill
\begin{minipage}{.48\textwidth}
        \includegraphics[width=\textwidth]{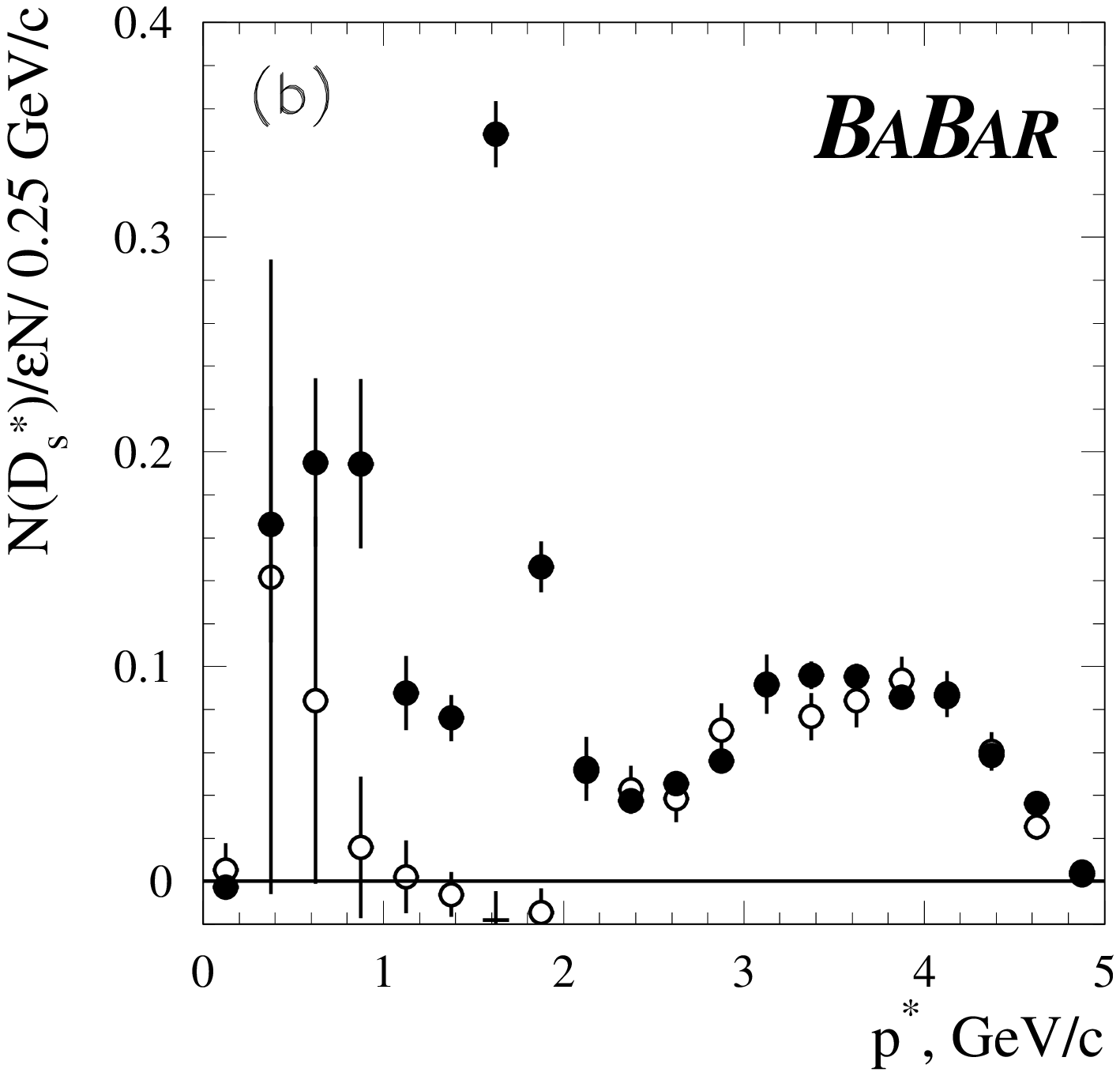}
\end{minipage}
\caption[.]
{The (a) \Ds and (b) \Dspstar efficiency-corrected 
momentum spectra for on-resonance data (solid circles)
and for scaled off-resonance data (open circles).}
\label{fig:nev}
\end{figure}
\end{center}

The efficiency obtained from Monte Carlo with generic \BB and \ccbar events  varies as a function of 
the \Dsps momentum ($p^*$) in the \FourS rest frame and 
ranges from 20\% when the \Ds is
at rest to 40\% for $p^* = 5 \gevc$, and from 5\% to 20\% for \Dspstar. 
The number of reconstructed \Ds and \Dspstar is corrected bin-by-bin for the efficiency.
The efficiency-corrected number of \Ds and \Dspstar
as a function of their momentum in the \FourS rest frame
is shown in Fig.~\ref{fig:nev}.

\section{Branching fractions}
\par
The \Ds and \Dspstar cross sections for production from 
the \qqbar continuum are obtained by 
%counting the number of events using 
integrating the spectrum obtained from
the off-resonance data. 
This gives the preliminary results
%\begin{equation}
%\label{math:ccbarDsX}
$$\sigma (e^+e^-\rightarrow D^{\pm}_sX)\times\BR(\Dsphipi) = 
7.55\pm0.20\pm0.34\ \pb$$
%\end{equation}
and
%\begin{equation}
%\label{math:ccbarDsstX}
$$\sigma (e^+e^-\rightarrow D^{*\pm}_sX)\times\BR(\Dsphipi) = 
5.79\pm0.66\pm0.50\ \pb.$$
%\end{equation}

The off-resonance data are scaled according to 
the luminosity ratio and then subtracted bin-by-bin from 
the on-resonance data 
in order to find the \Dsps momentum spectra from \B meson decays.
It is important to note that, with this method, the result is independent of
any assumption about the shape of the fragmentation function,
%================================
and most of the systematic errors due to the background 
parameterization cancel.
%================================
%Counting the number of events
Integrating the spectrum
after continuum subtraction
gives a total \Ds yield from \B meson 
decays of 87711$\pm$1485. This corresponds to 
the inclusive preliminary branching fraction
%\begin{equation}
%\label{math:BToDsX}
$$\BR(B\rightarrow D_s^+X) = \Biggl[(10.93\pm0.19\pm0.58)\times 
\frac{3.6\pm0.9\%}{\BR(D_s^+\rightarrow\phi\pi^+)}\Biggr]\%.$$
%\end{equation}
The total \Dspstar yield from \B meson 
decays is 60047$\pm$6201 events, leading to the 
inclusive preliminary branching fraction
%\begin{equation}
%\label{math:BToDsstarX}
$$\BR(B\rightarrow D_s^{*+}X) = \Biggl[(7.94\pm0.82\pm0.72)\times 
\frac{3.6\pm0.9\%}{\BR(D_s^+\rightarrow\phi\pi^+)}\Biggr]\%.$$
%\end{equation}
%
In the results above, the first error is statistical, 
the second is 
the systematic error and the third error, which is dominant, 
is due to the uncertainty in the \Dsphipi branching 
fraction~\cite{pdg}. The various contributions
to the systematic error are listed in Table~\ref{tab:syst}.
One of the dominant systematic errors is the 3.6\% total uncertainty 
due to our knowledge of the tracking efficiency (1.2\% per track
for the decay chain \Dsphipi, $\phi\rightarrow { K^+K^-}$).

\begin{table}
\begin{center}
\caption{ Systematic errors for $\BR(B\rightarrow \Dsps X)$}
\label{tab:syst}
\begin{tabular}{lcc}
\hline
\hline
Source  & \multicolumn{2}{c}{Fractional Error on \BR\ (\%)}      \\ \hline
        &       $B\rightarrow \Ds X$\hspace{0.5cm}              &
        $B\rightarrow \Dspstar X$       \\ \hline
Signal shape                                    &0.5    &       3.0     \\
Background parameterization                     &0.4    &       4.2     \\
%Continuum subtraction                           &      &       1.2    \\
Monte Carlo statistics                          &2.5    &       4.2    \\
Bin width                                       &1.4    &       2.0     \\ 
\hline
Total for \Ds yield                             &2.9    &       7.0    \\
Number of \BB events                            &1.6    &       1.6     \\
$\BR(\phi\rightarrow { K^+K^-})$        &1.6    &       1.6   \\
Particle id  efficiency                         &1.0    &       1.0   \\
Tracking efficiency                             &3.6    &       3.6    \\ 
$\BR(\Dsgamma)$                                 &       &       2.7    \\
Photon efficiency                               &       &       1.3     \\
$\pi^0$ veto                                    &       &       2.7     \\ 
\hline
Total systematic error                          &5.3    &       9.0   \\ 
\hline \hline
\end{tabular}
\end{center}
\end{table}

\section{\boldmath Fits to \Dsps momentum spectra}
\par
In the \FourS rest frame, two-body \B decays 
produce \Dsps mesons with a flat momentum spectrum over a 300\mevc\ wide
range.
In \B decays, the \Dsps momentum spectrum is essentially governed by the 
production of direct \Dsps.
Other $c\overline{s}$ states such as $D_{s1}(2536)$ and $D_{s2}^*(2573)$
primarily decay to $D^{(*)}K$. 
Because \Dspstar decays to $\Ds\gamma$ or $\Ds\pi^0$, the \Ds momentum 
distribution is slightly broader and shifted 
downward compared to direct production from $B\rightarrow \Ds X$.

In fitting the observed momentum spectra,
three different sources of \Dsps mesons in \B decays are considered:
\begin{enumerate}
\renewcommand{\labelenumi}
                {(\arabic{enumi})}
\item $B \rightarrow \Dsps \Dbar^{(*)}$ decays.
The relative branching fractions of the individual channels
can be taken either from existing measurements~\cite{cleo:dsinc}
or from predictions assuming factorization~\cite{th_BSW, th_rosner, th_neubert}.  
The fit is performed for both cases, with the assumption
$f_{\Dspstar}=f_{\Ds}$ for the theoretical models, where   
$f_{\Dsps}$ are the \Dsps decay constants.
\item $B\rightarrow \Dsps \Dbar^{**}$ decays. 
The contributions from $B\rightarrow \Dsps \Dbar_0^*(j=1/2)$, 
$B\rightarrow \Dsps \Dbar_1(2420)$, 
$B\rightarrow \Dsps \Dbar_1(j=1/2)$ and 
$B\rightarrow \Dsps \Dbar_2^*(2460)$ 
are included in this source.
\item Three-body ${ B\rightarrow \Dsps \Dbar^{(*)} \pi/\rho/\omega }$ decays.
Since little is known on these decays, 
they are attributed an equal weight and the momentum 
distributions are generated according
to phase space.
\end{enumerate}

\begin{center}
\begin{figure}
\begin{minipage}{.48\textwidth}
        \includegraphics[width=\textwidth]{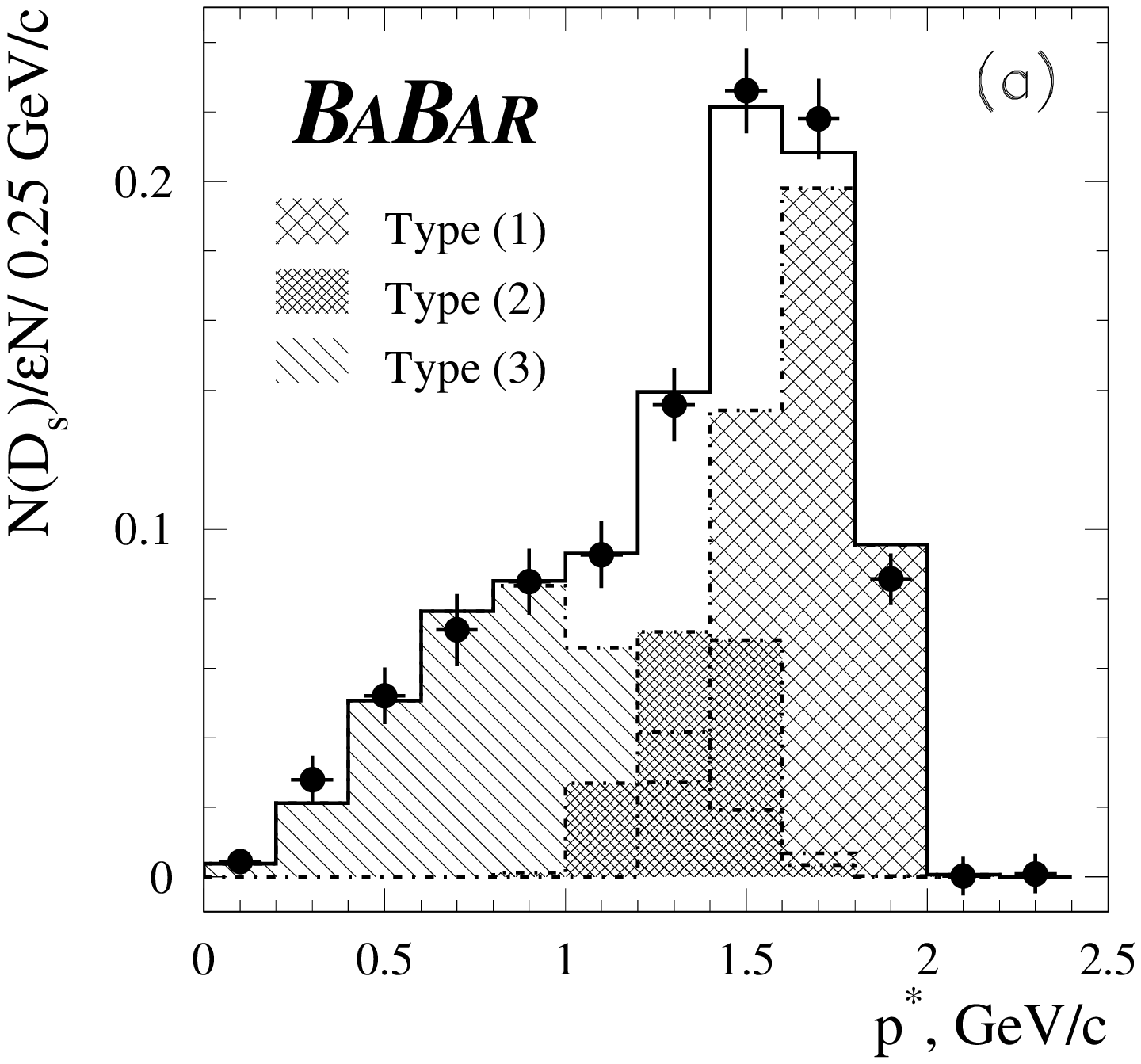}
\end{minipage}
\hfill
\begin{minipage}{.48\textwidth}
        \includegraphics[width=\textwidth]{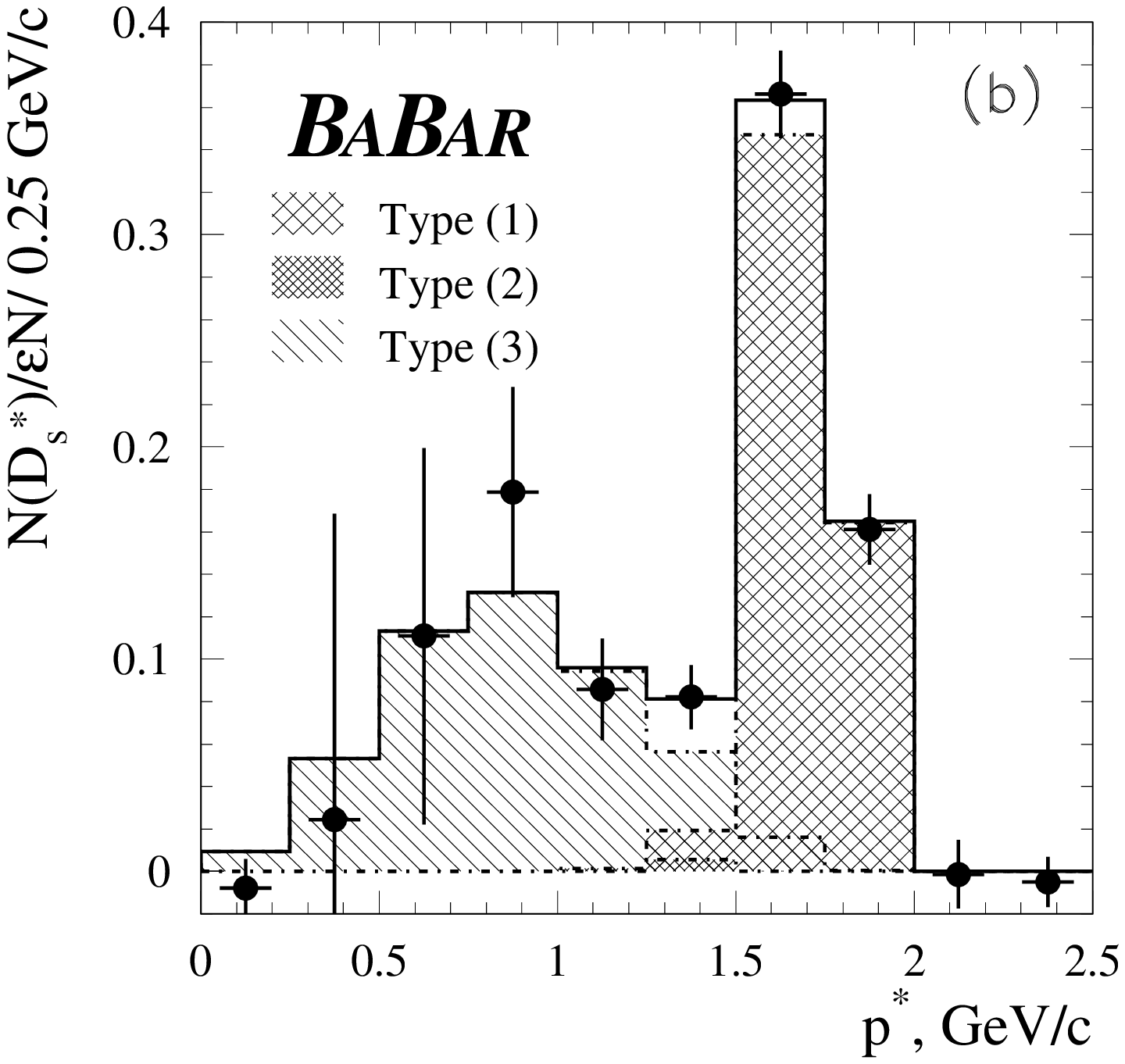}
\end{minipage}
\caption[.]
{
The fit result for (a) \Ds and (b) \Dsps momentum spectra. 
The data are dots with error bars, the histograms are the components 
of the fit function described in the text. Type (1) is
$B\rightarrow \Dsps \Dbar^{(*)}$, Type (2) is
$B\rightarrow \Dsps \Dbar^{**}$ and Type (3) is
$B\rightarrow \Dsps \Dbar^{(*)} \pi/\rho/\omega$.
The solid histogram is the sum of three components.
}
\label{fig:fit}
\end{figure}
\end{center}

As a result of a chi-squared fit of \Ds\ momentum spectrum with these components, 
the ratio of two-body modes to the total inclusive rate is
determined to be
$$ {\Sigma\BR(B\rightarrow \Dsps \Dbar^{(*)}) 
\over \BR(B\rightarrow \Ds X)} =
(46.4 \pm 1.9 \pm 0.6) \%, $$
where the first error is statistical and the second is due to model uncertainty.
This last is obtained from the variation of the fit result with different 
individual contributions from the modes included with each of the three sources
of \Ds mesons, as discussed below.
From the fit to the \Dspstar\ momentum spectrum, we find
$$ {\Sigma\BR(B\rightarrow \Dspstar \Dbar^{(*)}) 
\over \BR(B\rightarrow \Dspstar X) } =
(53.3 \pm 4.5 \pm 1.6 \pm 2.1) \%.$$
%for \ensuremath{ B\rightarrow D_s^{*+} \Dbar^{(*)} }.
where the first error is statistical,
the second error represents the systematic uncertainty
due to the background parameterization
(negligible for $B\rightarrow \Dsps \Dbar^{(*)}$), and
the third error is due to model uncertainty 
obtained as for 
$B\rightarrow D_s^{(*)+} \Dbar^{(*)}$.

%The second error in the fraction of $B\rightarrow \Dspstar \Dbar^{(*)}$
%shows the systematic uncertainties defined by background parameterization.
%It was shown that this error is significantly lower than statistical error
%for the fraction of $B\rightarrow \Dsps \Dbar^{(*)}$. It is interesting
%to note that other sources of the systematic errors presented in 
%Table~\ref{tab:syst}
%are correlated and do not give the contribution to these numbers.

The fit is performed under different assumptions for the relative 
contributions of the modes in source (1), varied
according to the theoretical predictions and measurements. 
Different weights of
${ B\rightarrow D_s^+ \Dbar^{**}}$ 
and ${ B\rightarrow D_s^{*+} \Dbar^{**}}$,
as well as different relative branching fractions 
of the four modes within source (2), are also used.
In source (3), two cases are considered:
either $B\rightarrow D_s^{(*)} \Dbar^{(*)} \pi$ or 
$B\rightarrow D_s^{(*)} \Dbar^{(*)} \rho/\omega$ 
is assumed to be dominant. The best $\chi^2$ for the fit to the
inclusive \Dspstar momentum spectrum is obtained when the contribution from
$B\rightarrow D_s^{(*)} \Dbar^{(*)} \rho/\omega$ 
is dominant compared to
$B\rightarrow D_s^{(*)} \Dbar^{(*)} \pi$.
The results of the fits to the \Dsps momentum spectra are shown in 
Fig.~\ref{fig:fit} for one of the assumptions.
% Which???

Using the fit results and the relative rates 
for $B\rightarrow D_s^{(*)} \Dbar^{(*)}$ we find the preliminary results
%\begin{equation}
%\label{math:SumDsD}
$$\Sigma\BR(B\rightarrow \Dsps \Dbar^{(*)}) = 
(5.07\pm0.09\pm0.34\pm1.27)\%,$$
%\end{equation}
%\begin{equation}
%\label{math:SumDsstD_ncorr}
$$\Sigma\BR(B\rightarrow D_s^{*+} \Dbar^{(*)}) = 
(4.07\pm0.42\pm0.53\pm1.02)\%.$$
%\end{equation}
where the errors from the fits to the momentum spectra are added in 
quadrature with the systematic error due to the \Dsphipi
branching fraction uncertainty.

%==================================================================
%
%       SUMMARY
%
%==================================================================
\section{Summary}
\par
In summary, preliminary branching fractions for inclusive 
$B\rightarrow\Dsps X$ production have been determined to be
$\BR(B\rightarrow D_s^+X) = (10.93\pm0.19\pm0.58\pm2.73)\%$ and
$\BR(B\rightarrow D_s^{*+}X) = (7.94\pm0.82\pm0.72\pm1.99)\%$.
The \Dsps cross sections from \qqbar continuum events
at about 40\mev below \FourS mass are
$\sigma (e^+e^-\rightarrow D^{\pm}_s  X)\times\BR(\Dsphipi) =
7.55\pm0.20\pm0.34\ \pb$ and
$\sigma (e^+e^-\rightarrow D^{*\pm}_s X)\times\BR(\Dsphipi) =
5.79\pm0.66\pm0.50\ \pb$.
%The measurements of $B\rightarrow\Dspstar X$ and its 
%cross section from continuum events
%have been obtained for the first time. 
Our results for \Ds are in agreement with previous
measurements~\cite{cleo:dsinc, argus}, although
with considerable improvement in accuracy. 
In contrast to previous measurements, 
our results do not rely on any assumptions concerning the shape of
the fragmentation function.

From a fit to the \Dsps momentum spectra,
preliminary results have been obtained for the fraction of all two-body 
$B\rightarrow \Dsps \Dbar^{(*)}$ decays
relative to the total inclusive \Ds yield ($46.4\pm1.9\pm0.6$)\%
and for all
$B\rightarrow D_s^{*+} \Dbar^{(*)}$ decays 
relative to the total inclusive \Dspstar yield
($53.3\pm4.5\pm1.6\pm2.1$)\%, where
the last error includes the model dependence. Combining these
results gives
$\Sigma\BR(B\rightarrow \Dsps \Dbar^{(*)}) = (5.07\pm0.09\pm0.34\pm1.27)\%$
and 
$\Sigma\BR(B\rightarrow \Dspstar \Dbar^{(*)}) = (4.07\pm0.42\pm0.53\pm1.02)\%$

%====================================================================

\section{Acknowledgements}
We are grateful for the 
extraordinary contributions of our \pep2\ colleagues in
achieving the excellent luminosity and machine conditions
that have made this work possible.
The collaborating institutions wish to thank 
SLAC for its support and the kind hospitality extended to them. 
This work is supported by the
US Department of Energy
and National Science Foundation, the
Natural Sciences and Engineering Research Council (Canada),
Institute of High Energy Physics (China), the
Commissariat \`a l'Energie Atomique and
Institut National de Physique Nucl\'eaire et de Physique des Particules
(France), the
Bundesministerium f\"ur Bildung und Forschung
(Germany), the
Istituto Nazionale di Fisica Nucleare (Italy),
the Research Council of Norway, the
Ministry of Science and Technology of the Russian Federation, and the
Particle Physics and Astronomy Research Council (United Kingdom). 
Individuals have received support from the Swiss 
National Science Foundation, the A. P. Sloan Foundation, 
the Research Corporation,
and the Alexander von Humboldt Foundation.


\begin{thebibliography}{99}
\bibitem{CBfunc}
The $\Delta M$ distribution for the $\Ds\gamma$ signal is fit
with the Crystal Ball function
  \[
  f(x) = N \cdot \left\{
  \begin{array}{ll}
    \exp (-\frac{(x - \bar {x})^2}{2 \; \sigma^2})  &;  
( x - \bar {x})/\sigma >   \alpha \\
    A \times (B - \frac{x - \bar {x}}{\sigma})^{-n} &;  
( x - \bar {x})/\sigma \leq \alpha \\
  \end{array} \right. \]
  where $A \equiv \left ( \frac{n}{|\alpha|} \right ) ^n 
\times \exp(-|\alpha|^2/2)$ and
  $B \equiv \frac{n}{|\alpha|} - |\alpha|$.
  $N$ is a normalization factor, $\bar x$ and $\sigma$ are the peak position 
and width
  of the Gaussian portion of the function, $\alpha$ is the point at which the 
function
  changes to the power function and $n$ is the exponent of the power function.
  $A$ and $B$ are defined such as to maintain the continuity of the function 
and it first
  derivative at $\alpha$.
  More details can be found in D.\ Antreasyan, Crystal Ball Note 321 (1983).
\end{thebibliography}
\end{document}